\documentclass[10pt]{ismd08}
\usepackage{graphicx}
\usepackage{cite,./mcite}

\begin{document}
\title{Discovery potential at the LHC: channels relevant for SM Higgs}
\author{I. Tsukerman on behalf of ATLAS and CMS collaborations}
\institute{ITEP, Moscow, Russia}
\maketitle
\begin{abstract}
The discovery potential of Standard Model Higgs searches at the LHC at 14 TeV center-of-mass energy 
is reviewed. Decay channels such as $H\rightarrow\gamma\gamma$, $H\rightarrow~ZZ^{\star}\rightarrow~4\ell$, 
$H\rightarrow~WW^{\star}$ and $H\rightarrow~\tau\tau$ are considered. Results are based on
the most recent full GEANT-based simulations performed by the ATLAS and CMS experiments.
\end{abstract}
\section{Introduction}
The primary objective of the Large Hadron Collider (LHC) at CERN is to 
study the origin of electroweak symmetry breaking.
Within  the Standard Model (SM), the Higgs mechanism \cite{higgs1,*higgs2,*higgs3} is invoked
to explain this breaking and the Higgs boson remains the only particle that has not been
discovered so far. The direct search at the $e^{+}e^{-}$ collider LEP has led to a lower
bound on its mass of 114.4 GeV at 95\% C.L. \cite{lep}.
In addition, high precision electroweak data constrain the mass of the 
Higgs boson via their sensitivity to loop corrections. The upper limit 
is $m(H)\leq$ 185 GeV at 95\% C.L. provided the LEP result is also used in the
determination of this limit \cite{lepewwg}. At last, combined preliminary results
from the Tevatron experiments CDF and D0 based on 3 $fb^{-1}$ accumulated data at 
1.8 TeV 
lead to a 95\% C.L. exclusion of a Higgs boson with $m(H)$=170 GeV \cite{fnal}.  
Both ATLAS and CMS experiments at the LHC, scheduled for proton-proton collision
data taking in summer 2009, have been designed to search for the Higgs boson over a
wide mass range \cite{atlas,*cms}. In these proceedings the sensitivity for each
experiment to discover or exclude the SM Higgs boson as well as recent developments
that have enhanced this sensitivity are summarized.   
\section{SM Higgs production at the LHC}
Theoretical predictions for NLO SM Higgs production cross-sections at 14 TeV energy as a function of $m(H)$
\cite{theorh} are shown in Fig.\ref{sigma} (left). The dominant production mechanism, which proceeds via
a top-quark loop, is gluon-gluon fusion ($gg\rightarrow~H$). It gives rise to 20--40 pb
SM Higgs cross section in the mass range between 114 and 185 GeV.
The vector-boson fusion (VBF) process ($qq\rightarrow~qqH$) has a factor of eight smaller 
cross section. However, in this case, the Higgs boson is accompanied by two energetic jets 
going mainly into the forward directions. Usually they have large pseudorapidity gap in-between.
In addition, there is no colour flow between these {\it tagging} jets which allows for use
of {\it a central jet veto} to reduce backgrounds. 
$q\bar{q}\rightarrow HW$, $q\bar{q}\rightarrow HZ$ and $gg,q\bar{q}\rightarrow t\bar{t}H$ processes
have smaller cross sections. 
\section{SM Higgs discovery final states}
\begin{figure*}[t]
\centering
\includegraphics[width=58mm]{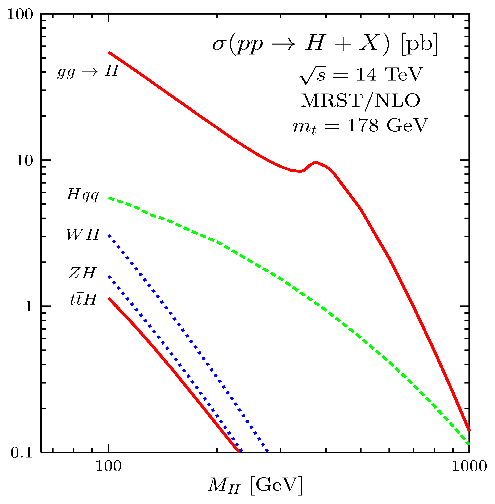}
\includegraphics[width=59mm]{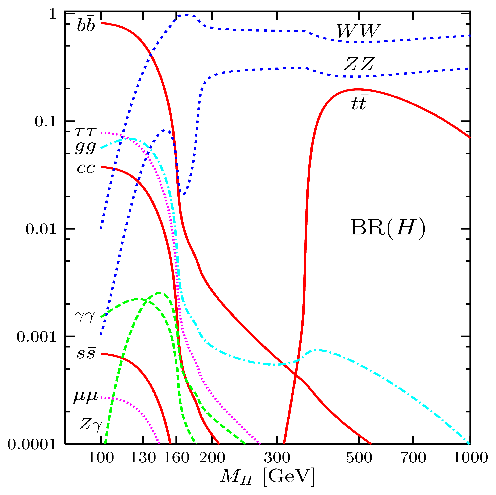}
\caption {
Left: Theoretical predictions for SM Higgs boson production cross sections at LHC energies.  
Right: Theoretical predictions for SM Higgs boson decay branching ratios.}
\label{sigma}
\end{figure*}

The SM Higgs boson is predicted to have many decay channels 
with branching ratios which strongly depend on its mass
(Fig.\ref{sigma} (right)). The evaluation of the search sensitivity
of the various channels should take into account the cross-sections
of the relevant backgrounds.

At low Higgs mass the dominant decay mode is through $b\bar{b}$. However,
due to the enormous QCD backgrounds this channel is not good for the SM Higgs discovery.
The $\gamma\gamma$ final state, which appears when the Higgs
decays via bottom, top and $W$-loops, has a small branching fraction. However,
excellent diphoton invariant-mass resolution and $\gamma$/jet separation can make
this mode one of the best discovery channels. 
$H\rightarrow \tau\tau$ has a sizeable rate and should be visible 
with good purity via the VBF Higgs production mode.

If the Higgs mass is larger, the $H\rightarrow~WW^{\star}$ final states are powerful
as well as the mode $H\rightarrow~ZZ^{\star}\rightarrow~4\ell$. In the last case, 
the resulting branching ratio is small but the signal is easy to trigger on 
and allows for full reconstruction of the Higgs mass. 

Both ATLAS and CMS Collaborations have performed 
extensive GEANT-based Monte Carlo \cite{gean1,*gean2} studies
with full simulation and reconstruction to determine the experimental
viability of many Higgs decay channels. Results of the
recent studies \cite{atlcsc,*cmstdr} for the most attractive signatures, namely
$H\rightarrow \gamma\gamma$, $H\rightarrow ZZ^{\star} \rightarrow 4\ell$, $H\rightarrow WW^{\star}$ and
VBF $H\rightarrow \tau\tau$ are summarized below.\footnote{Another summaries of
SM Higgs searches were presented at this year conferences, see, e.g.
Ref.\cite{goncalo,*vickey,*nisati}.}   
\subsection{$H\rightarrow \gamma\gamma$}
Despite a only 0.2\% branching ratio in the Higgs mass region 120--140 GeV,
$H\rightarrow \gamma\gamma$ remains a promising channel as the signal signature 
is very clean. Irreducible backgrounds come from continuum production of diphotons, 
$q\bar{q}, gg\rightarrow \gamma\gamma$. Reducible backgrounds are mostly
due to $\gamma$-jet and jet-jet events, where one or more jets are misidentified
as photons. Studies performed by both the ATLAS and CMS experiments consider the
signal and backgrounds at NLO level. Thanks to a very good
electromagnetic energy resolution, with a simple cut-based analysis 
for an integrated luminosity of 30 $fb^{-1}$ one can obtain a significance
above 5$\sigma$ in the CMS experiment for the mass 
range 115--140 GeV (Fig.\ref{sign} (top left)).  
Having a worse energy resolution, ATLAS nevertheless can reach almost
the same significance as the high-granularity electromagnetic calorimeter
with longitudinal samplings is capable of determining the primary vertex with 
great precision. Both experiments have looked beyond a simple cut-based
analysis and enhanced the signal significance by 30--50\% (Fig.\ref{sign} (top left)). 
\subsection{$H\rightarrow ZZ^{\star} \rightarrow 4\ell$}
The ``golden'' channels (4$\mu$, 2$e$2$\mu$ and 4$e$ final states of $ZZ^{\star}$ decays) 
are expected to be good for discovery in a wide mass range 
(except $m(H)\leq$130 GeV and $m(H)\approx 2m_{W}$). 
The dominant background is the $ZZ^{\star}$-continuum with 
smaller contributions from $Zb\bar{b}$ and $t\bar{t}$ processes.
Through the use of impact parameter and lepton isolation requirements the latter
two (which are important only at low $m(H)$) can be significantly reduced. Simulations 
of the signal and the $q\bar{q}\rightarrow ZZ^{\star}$ backgrounds were made up to the NLO level.
An additional 20--30\% contribution from the $gg \rightarrow ZZ^{\star}$ process
was also taken into account. 
A 5$\sigma$ discovery in $H\rightarrow ZZ^{\star} \rightarrow 4\ell$ mode
is possible in much of the allowed $m(H)$ space with less than
30 $fb^{-1}$ of integrated LHC luminosity (Fig.\ref{sign} (top right)).
\subsection{$H\rightarrow WW^{\star}$}
$H\rightarrow WW^{\star}$ is the main search channel
in the Higgs mass range $2m_{W}\leq m(H) \leq 2m(Z)$ due to a very large
$H\rightarrow WW$ branching ratio (Fig.\ref{sigma} (left)). This mode is also good at lower masses
(down to $m(H)\approx$ 130 GeV) and at high $m(H)$. Two different
final states are considered: $\ell\nu\ell\nu$ and $\ell\nu qq$. 
Unlike the $H\rightarrow ZZ \rightarrow 4\ell$ and $H\rightarrow \gamma\gamma$ channels, 
full mass reconstruction is not possible therefore an
accurate background estimate is critical. 
The dominant background for this analysis is $q\bar{q},gg\rightarrow WW^{\star}$-production in the case of
$H$ + 0 jets signal. This background
can be suppressed by exploiting the spin correlation between the two final state leptons.
For $H$ + 2 jets, where the contribution from $qq\rightarrow qqH$ process is the most important,
$t\bar{t}$-production is the main background which can be reduced 
by the {\it forward jet tagging} and {\it central jet veto} requirements.
NLO-level studies (with systematics included) have shown that less than 2 $fb^{-1}$ integrated
luminosity would be sufficient for a 5$\sigma$ discovery of the SM Higgs with $m(H)=$160--170 GeV.
Let us note that using the VBF $H\rightarrow WW^{\star}\rightarrow e\nu\mu\nu$ mode alone, 
ATLAS is able to observe this particle with 10 $fb^{-1}$ of integrated luminosity
provided 150 GeV $\leq m(H)\leq$ 180 GeV (Fig.\ref{sign} (bottom left)).
\subsection{VBF $H\rightarrow \tau\tau$}
\begin{figure*}[bhtp]
\centering
\includegraphics[width=59mm]{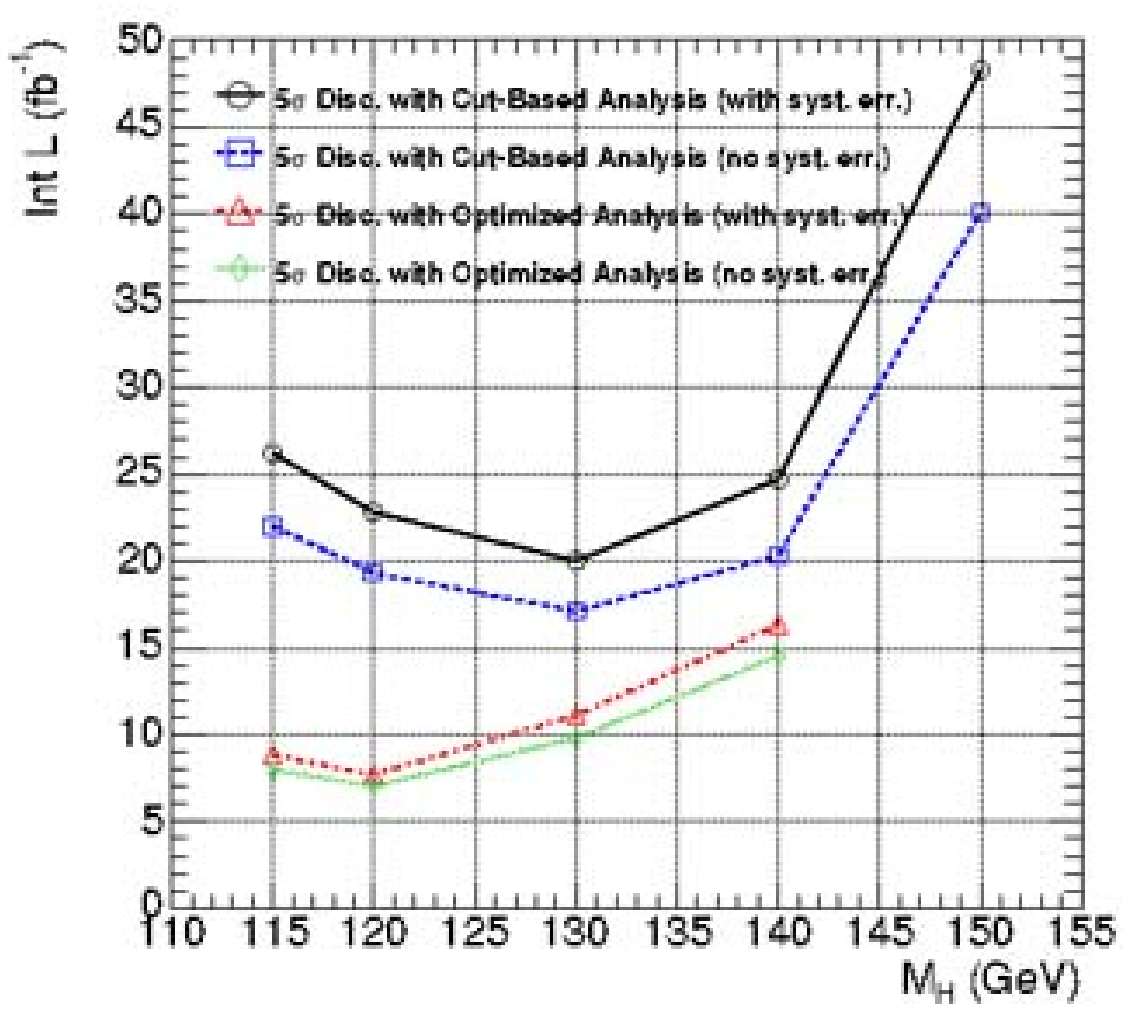}
\includegraphics[width=69mm]{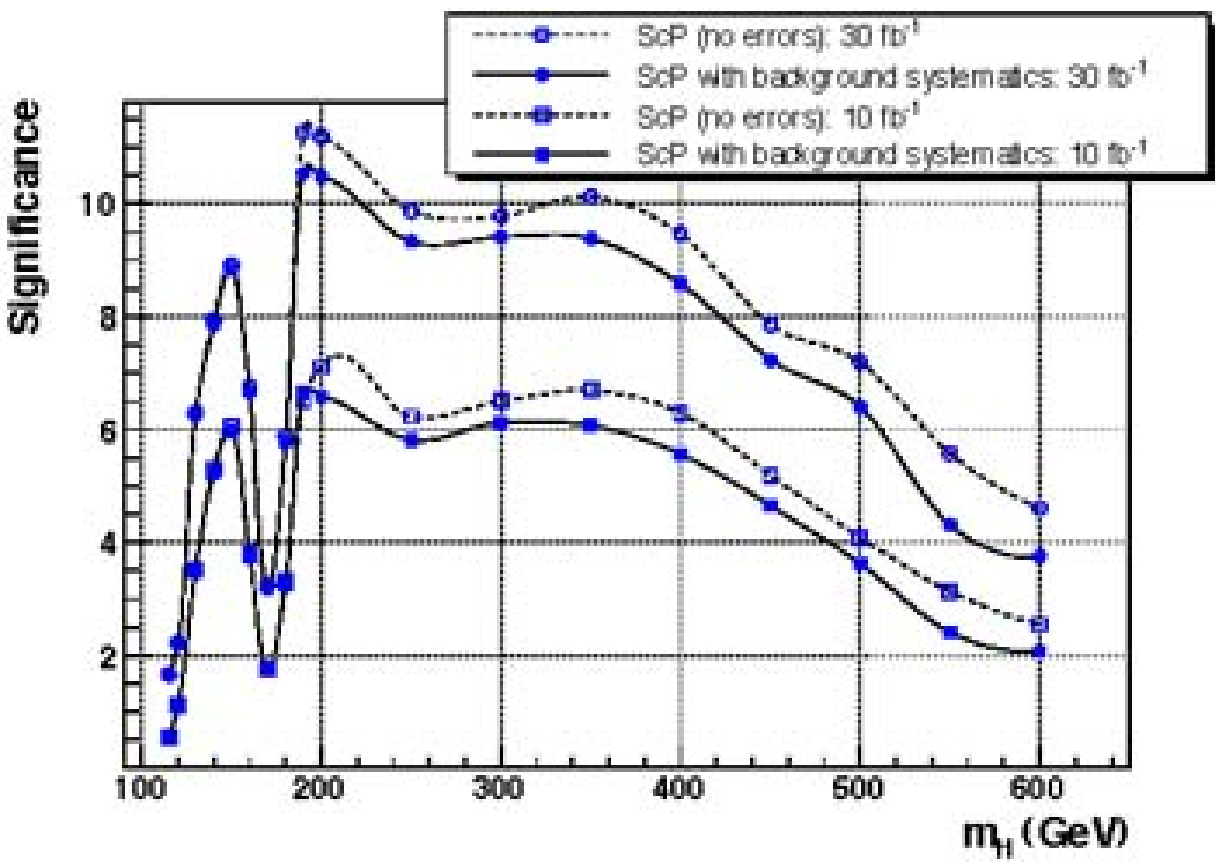}
\includegraphics[width=65mm]{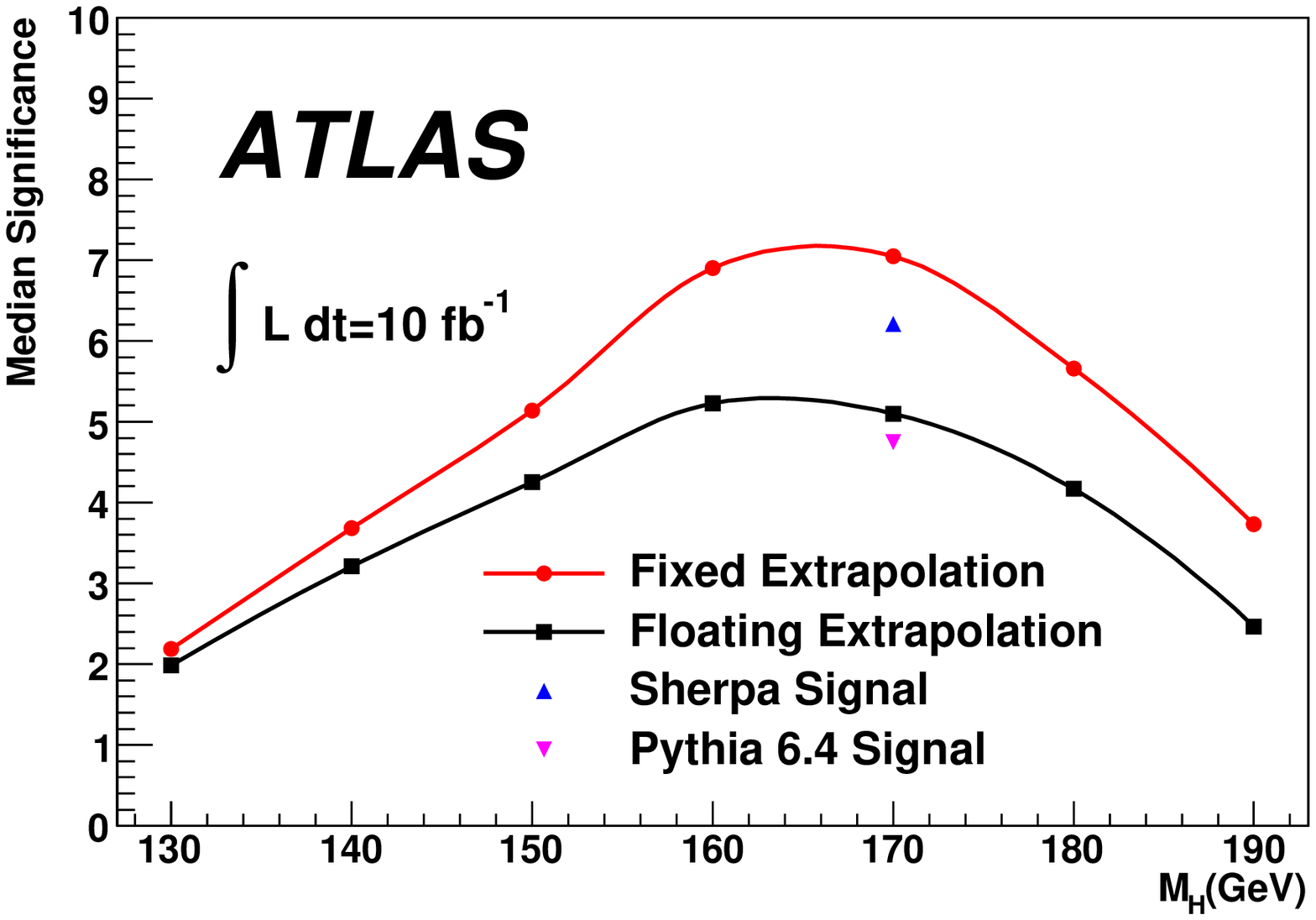}
\includegraphics[width=49mm]{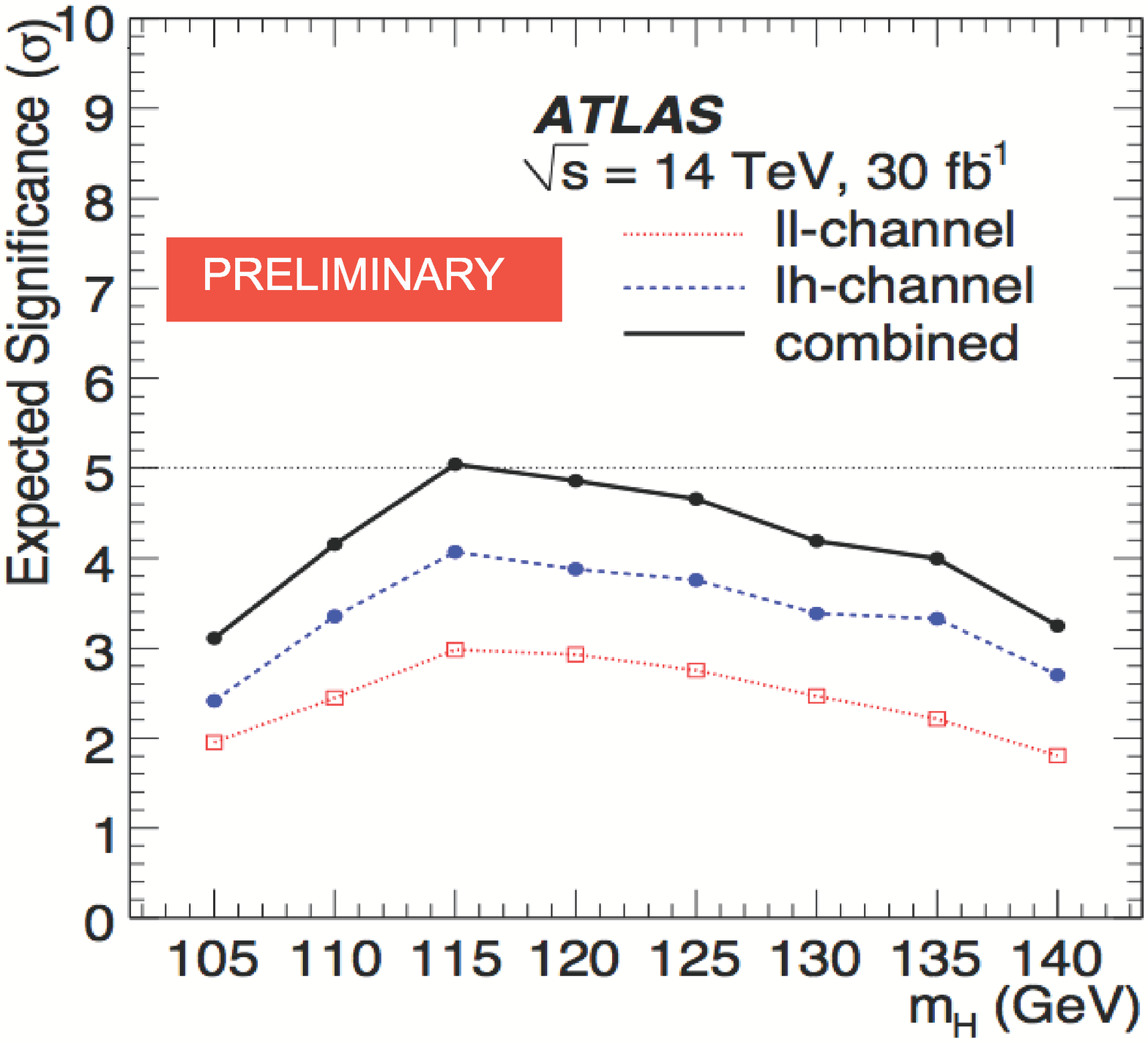}
\vspace*{-4mm}
\caption {SM Higgs discovery potential for specific decay modes.
Top left: CMS, $H\rightarrow \gamma\gamma$.
Top right: CMS,  $H\rightarrow ZZ^{\star}\rightarrow 4\ell$.
Bottom left: ATLAS, VBF $H\rightarrow WW^{\star}\rightarrow e\nu\mu\nu$.
Bottom right: ATLAS, VBF $H\rightarrow \tau\tau$.}
\label{sign}
\end{figure*}

In gluon-fusion production mode the $H\rightarrow \tau\tau$ channel is not promising
due to large backgrounds. However, one can consider the $qq\rightarrow qqH$ process 
which helps to reduce contributions
coming mainly from the $Z/\gamma^{\star}\rightarrow \tau\tau$ + jets and 
$t\bar{t}$ processes.  
Data-driven methods for understanding the dominant backgrounds 
have been investigated. Three final states of $\tau$ decays are considered: lepton-lepton, lepton-hadron 
and also hadron-hadron. Despite the presence of neutrinos,
mass reconstruction can be done via the collinear approximation
where $\tau$ decay daughters are assumed to go in the same directions
as their parents. The resolution on the reconstructed mass ($\sim$ 10 GeV)
is mainly affected by the missing transverse energy resolution. 
Simulations performed by ATLAS and CMS have shown that the
combination of the lepton-lepton and lepton-hadron channels should allow for a
5$\sigma$ measurement with 30 $fb^{-1}$ LHC luminosity in the
range 115 GeV $\leq m(H) \leq$ 125 GeV (Fig.\ref{sign} (bottom right)). 
\section{Summary of SM Higgs discovery potential}
\begin{figure*}[bhtp]
\centering
\includegraphics[width=74mm]{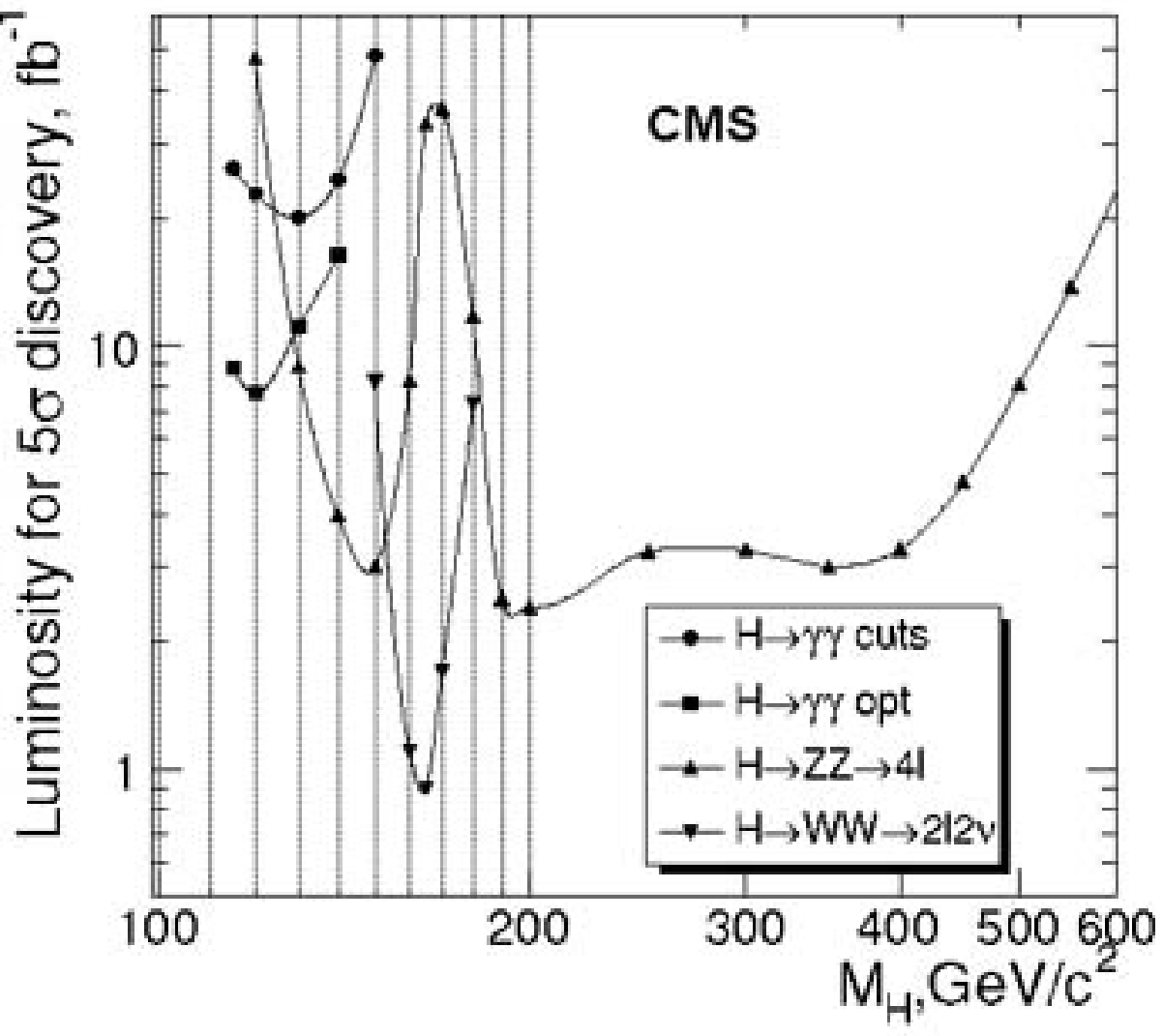}
\includegraphics[width=73mm]{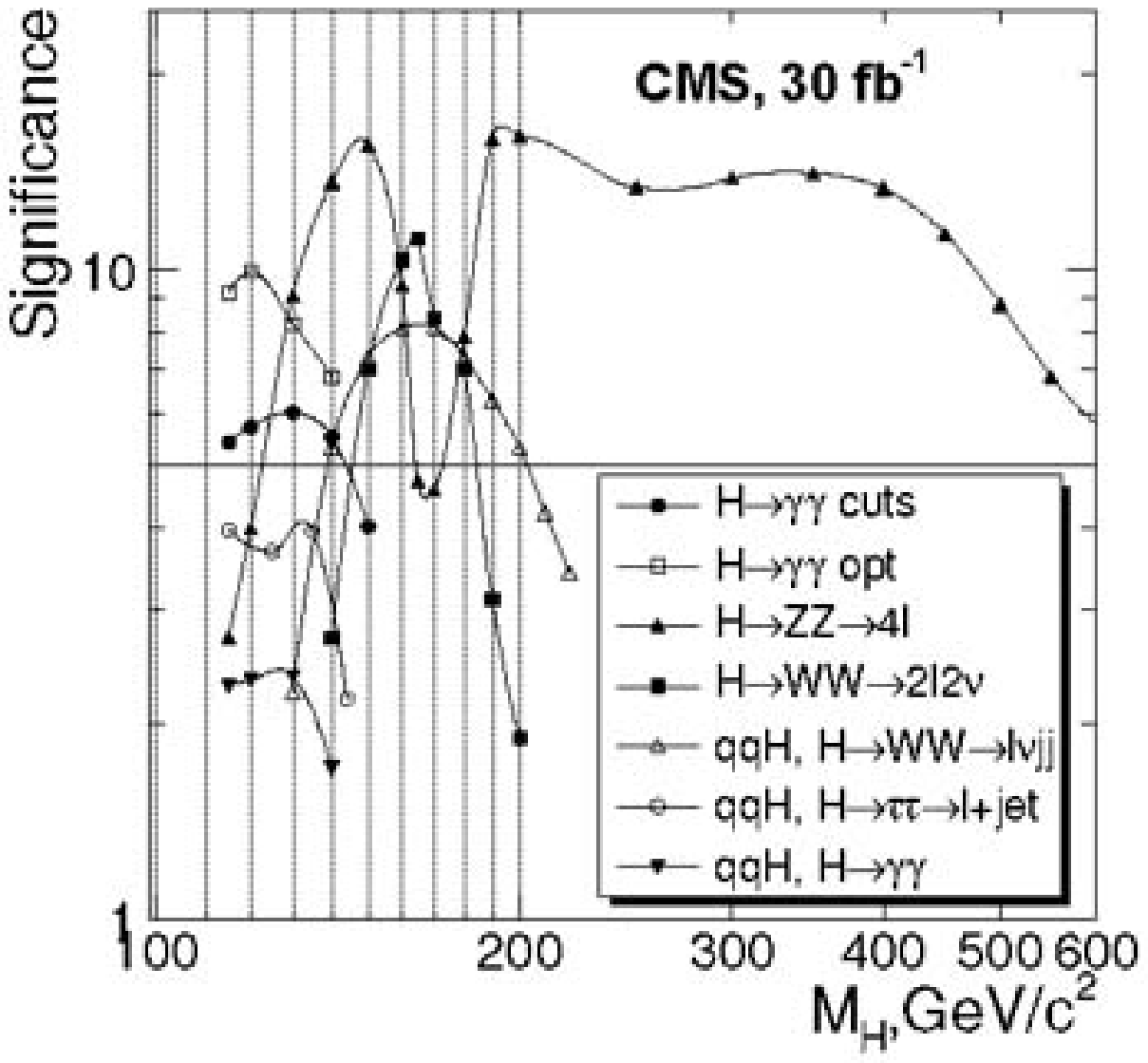}
\caption {Left: Integrated luminosity needed for 5$\sigma$ discovery of the
inclusive Higgs boson production pp$\rightarrow H$ + X with the Higgs boson decay modes
$H\rightarrow \gamma\gamma$, $H\rightarrow ZZ^{\star} \rightarrow 4\ell$ and
$H\rightarrow WW^{\star}\rightarrow \ell\nu\ell\nu$ in the CMS experiment.
Right: The signal significance as a function of the Higgs boson mass for
30 $fb^{-1}$ of the integrated luminosity for the different Higgs boson production and
decay channels in the CMS experiment.}
\label{cmsd1}
\end{figure*}

Figure \ref{cmsd1} (left) shows integrated luminosity needed for the 5$\sigma$ discovery
of the inclusive Higgs boson production with the decay modes 
$H\rightarrow \gamma\gamma$, $H\rightarrow ZZ^{\star} \rightarrow 4\ell$ and
$H\rightarrow WW^{\star}\rightarrow \ell\nu\ell\nu$ in the CMS experiment. 
In the most complicated region below $m(H)=$130 GeV, less than 
10 $fb^{-1}$ would be sufficient while in the range 155 GeV$\leq m(H)\leq$400 GeV
only 3 $fb^{-1}$ are required.
The signal significance as a function of 
the Higgs boson mass for 30 $fb^{-1}$ of integrated LHC luminosity 
for the different Higgs boson production and decay channels is shown
in Fig.\ref{cmsd1} (right). Here $H\rightarrow\tau\tau\rightarrow\ell\nu j\nu$ and
$H\rightarrow WW^{\star}\rightarrow \ell\nu jj$ final states are also
included. For $m(H)$ between 115 and 500 GeV 10$\sigma$ discovery can be reached.

\noindent In summary, 
one can conclude 
that with integrated LHC luminosity of $\sim$5 $fb^{-1}$
it is possible to discover SM Higgs boson provided its mass is above the
114 GeV limit obtained by LEP \cite{bla}. It is sufficient to accumulate 1 $fb^{-1}$
for a 95\% C.L. exclusion in the full allowed mass range.\footnote{These statements are based
on older simulations \cite{atltdr,*vbf,*cmstd}. 
However one would not expect major changes when new results will be included.}  
\section{Acknowledgments}
The author would like to thank the ATLAS and CMS Collaborations, in particular,
M. Duehrssen, L. Fayard, R. Goncalo, K. Jakobs, V. Khovanskiy, B. Mellado, B. Murray,
A. Nikitenko, A. Nisati, W. Quayle, T. Vickey and B. Wosiek.  
\begin{footnotesize}
\bibliographystyle{ismd08} 
{\raggedright
\bibliography{iitismd08new}
}
\end{footnotesize}

\end{document}